\documentclass[aps,prb,preprint,groupedaddress,showpacs,amsmath,amssymb]{revtex4}

\usepackage{graphicx}
\usepackage{dcolumn}
\usepackage{bm}
\usepackage{hyperref}

\begin{document}

\title{Coexistence of superconductivity and magnetism in  CaK(Fe$_{1-x}$Ni$_x$)$_4$As$_4$ as probed by $^{57}$Fe M\"ossbauer spectroscopy}

\author{Sergey L. Bud'ko$^{1,2}$, Vladimir G. Kogan$^{1}$, Ruslan Prozorov$^{1,2}$, William R. Meier$^{2}$, Mingyu Xu$^{1,2}$, and  Paul C. Canfield$^{1,2}$ }
\affiliation{$^{1}$Ames Laboratory, US DOE, Iowa State University, Ames, Iowa 50011, USA}
\affiliation{$^{2}$Department of Physics and Astronomy, Iowa State University, Ames, Iowa 50011, USA}

\date{\today}

\begin{abstract}

Temperature dependent $^{57}$Fe M\"ossbauer spectroscopy and specific heat measurements for  CaK(Fe$_{1-x}$Ni$_x$)$_4$As$_4$ with $x$ = 0, 0.017, 0.033, and 0.049 are presented. No magnetic hyperfine field (e.g. no static magnetic order) down to 5.5 K was detected for $x$ = 0 and 0.017 in agreement with the absence of any additional feature below superconducting transition temperature, $T_c$, in the specific heat  data. The evolution of magnetic hyperfine field with temperature was studied for $x$ = 0.033 and 0.049. The long-range magnetic order in these two compounds coexists with superconductivity.  The magnetic hyperfine field, $B_{hf}$,  (ordered magnetic moment) below $T_c$ in  CaK(Fe$_{0.967}$Ni$_{0.033}$)$_4$As$_4$  is continuously suppressed with the developing superconducting order parameter. The $B_{hf}(T)$ data for  CaK(Fe$_{0.967}$Ni$_{0.033}$)$_4$As$_4$, and  CaK(Fe$_{0.951}$Ni$_{0.049}$)$_4$As$_4$ can be described reasonably well by Machida's model for coexistence of itinerant spin density wave magnetism and superconductivity [K. Machida, J. Phys. Soc. Jpn. {\bf 50}, 2195 (1981)]. We demonstrate directly that superconductivity suppresses the spin density wave order parameter if the conditions are right, in agreement with the theoretical analysis.

\end{abstract}


\maketitle

\section{Introduction}

Co-existence and competition of superconductivity and magnetism has been of interest for condensed matter community for a long time. \cite{map76a,shr84a,buz84a,buz86a,kak88a,fis90a,can98a, mul01a,buz06a,car13a,wol15a} Whereas in the past superconductivity and magnetism were often originating from different subsystems (e. g. with magnetism coming from local moments of rare earth, R$^{3+}$ as in RRh$_4$B$_4$, RMo$_6$(S,Se)$_8$, RNi$_2$B$_2$C \cite{shr84a,buz84a,buz86a,kak88a,fis90a,can98a, mul01a,buz06a,car13a,wol15a,can98a,mul01a}), iron - based superconductors \cite{joh10a,can10a,ste11a,chu15a,all16a} offer the case of superconductivity and itinerant magnetism competing in the same, shared, electron subsystem. There is a commonly accepted understanding in these materials that one needs to sufficiently suppress magnetic  (spin density wave) order to induce and stabilize superconductivity. The competition between superconductivity and magnetism in iron - based superconductors (in particular, in  Ba(Fe$_{1-x}$T$_x$)$_2$As$_2$, T = Co, Ni) was observed as a reduction of the average static Fe moment below $T_c$ inferred from the integrated intensity of the antiferromagnetic reflection in neutron scattering  experiments. \cite{pra09a,chr09a,wan10a,luo12a}
$^{57}$Fe M\"ossbauer study in another member of the 122 family, Ba$_{0.75}$K$_{0.25}$Fe$_2$As$_2$, \cite{mun13a}  showed a decrease in the magnetic hyperfine field, but no change in the magnetic volume fraction below $T_c$, a result that was interpreted as an indication of the microscopic coexistence of magnetism and superconductivity.

Recently, several members of a new structure type in the family of iron-based superconductors, AeAFe$_4$As$_4$ (Ae = Ca, Sr, Eu; A = K, Rb, Cs), so-called 1144 superconductors,  were discovered. \cite{iyo16a,liu16a} These compounds are stoichiometric superconductors and do not require tuning by substitution or pressure to exhibit superconductivity. Successful growth and basic characterization of CaKFe$_4$As$_4$ single crystals \cite{mei16a,mei17a} opened the door for detailed studies of its superconducting and normal state properties. More importantly, it was followed by successful transition metal (Co and Ni) substitution for Fe in CaKFe$_4$As$_4$. \cite{mei18a}  As a result of this substitution, a new, spin-vortex-crystal magnetic phase \cite{oha17a} was stabilized in CaK(Fe$_{1-x}$T$_x$)$_4$As$_4$ (T = Co, Ni) and range of T - concentrations where superconductivity coexists with magnetism was outlined. \cite{mei18a} Bulk superconductivity in these samples was suggested by magnetic and transport measurements, as well as by the size of the jump in the specific heat  at $T_c$ (see Appendix A).

Given the unusual nature of the magnetic phase, availability of homogeneous single crystals, and accessible superconducting and magnetic ordering temperatures, these materials present a fertile playground to study competition between superconductivity and magnetism with microscopic, local probes. The elastic neutron scattering study of several of these compounds has been recently completed. \cite{kre18a} However, as discussed in Ref. [\onlinecite{mun13a}], Bragg intensities reflect the product of magnetic volume fraction and magnitude of magnetic moments, whereas $^{57}$Fe M\"ossbauer spectroscopy can address magnetic phase separation in the samples.

In this work we present temperature dependent $^{57}$Fe M\"ossbauer spectroscopy data on CaK(Fe$_{1-x}$Ni$_x$)$_4$As$_4$ samples with $x$ = 0.017, 0.033, and 0.049. Using these data we analyze coexistence and competition of superconductivity and magnetism in 1144 family, and refine $x - T$ phase diagram. The M\"ossbauer spectroscopy data will be compared with the results for pure, $x$ = 0, CaKFe$_4$As$_4$. \cite{bud17a}

\section{Experimental details}

Single crystals of CaK(Fe$_{1-x}$Ni$_x$)$_4$As$_4$ were grown out of a high-temperature solution rich in transition-metals and arsenic similar to the procedure used for the pure compound, see Refs. [\onlinecite{mei16a,mei17a,mei18a}] for further details. The Ni - composition in the samples was determined using  wavelength-dispersive x-ray spectroscopy. \cite{mei18a} The crystals were screened \cite{mei16a} to avoid possible contaminations by minority phases. M\"ossbauer spectroscopy measurements were performed using a SEE Co. conventional, constant acceleration type spectrometer in transmission geometry with a $^{57}$Co(Rh) source kept at room temperature. The absorbers were prepared as a mosaic of single crystals held on a VWR Weighting Paper disk by a small amount of Apiezon N grease. An effort was made to keep gaps between crystals to a minimum and the part of the disk not covered by crystals was coated with tungsten powder (Alfa Aesar 99.9\% metals basis).  The $c$ axis of the crystals in the mosaic was parallel to the M\"ossbauer $\gamma$ -  beam. The absorber was cooled to a desired temperature using a Janis model SHI-850-5 closed cycle refrigerator (with vibration damping). The driver velocity was calibrated using an $\alpha$ - Fe foil, and all isomer shifts (IS) are quoted relative to the $\alpha$ - Fe foil at room temperature.  A limited set of data for  CaK(Fe$_{0.951}$Ni$_{0.049}$)$_4$As$_4$ taken with different source and absorber was presented in Ref. [\onlinecite{mei18a}]. The M\"ossbauer spectra were fitted using the commercial software package MossWinn. \cite{kle16a}

\section{Results}

Subsets of M\"ossbauer spectra for CaK(Fe$_{1-x}$Ni$_x$)$_4$As$_4$ samples with $x$ = 0.017, 0.033, and 0.049 are shown in Fig. \ref{F1}.  For CaK(Fe$_{0.983}$Ni$_{0.017}$)$_4$As$_4$ [Fig. \ref{F1}(a)] the absorption lines are asymmetric, suggesting that each spectrum is a quadrupole split doublet with rather small value of the quadrupole splitting, QS. There are no extra features observed, confirming that the samples are single phase. For the spectrum taken at the base temperature, $T = 5.5$ K, there is no apparent broadening that could be associated with a hyperfine field at the $^{57}$Fe site, e.g. no evidence of a long range magnetic order, at least down to 5.5 K. All in all the M\"ossbauer spectra  for CaK(Fe$_{0.983}$Ni$_{0.017}$)$_4$As$_4$ are closely reminiscent of those for pure CaKFe$_4$As$_4$. \cite{bud17a}

 The evolution of the spectra on cooling for two other samples,  CaK(Fe$_{0.967}$Ni$_{0.033}$)$_4$As$_4$ and CaK(Fe$_{0.951}$Ni$_{0.049}$)$_4$As$_4$ [Figs. \ref{F1}(b),(c)], is very different. At high temperatures, in the paramagnetic state, the spectra are doublets that are very similar to those of CaKFe$_4$As$_4$ and CaK(Fe$_{0.983}$Ni$_{0.017}$)$_4$As$_4$. At low temperatures the spectra broaden and change their shape.  These low temperature data can be fit with a magnetic sextet. The full Hamiltonian approach ("Mixed $M + Q$  Static Hamiltonian (Mosaic)" model in the MossWinn \cite{kle16a} software package) was used to analyze these spectra. For $T \leq 40$ K (CaK(Fe$_{0.967}$Ni$_{0.033}$)$_4$As$_4$) and $T \leq 50$ K (CaK(Fe$_{0.951}$Ni$_{0.049}$)$_4$As$_4$) the the fits yield the angle $\theta$ between the directions of magnetic moments and $\gamma$ -  rays close to $90^\circ$, suggesting that the magnetic moments are in the $ab$ - plane, as has been argued in Ref. [\onlinecite{mei18a}].

The temperature dependence of the hyperfine field on $^{57}$Fe in CaK(Fe$_{0.967}$Ni$_{0.033}$)$_4$As$_4$ and CaK(Fe$_{0.951}$Ni$_{0.049}$)$_4$As$_4$  is shown in Fig. \ref{fig3}. For  CaK(Fe$_{0.951}$Ni$_{0.049}$)$_4$As$_4$ $B_{hf}$ increases smoothly on cooling below $\sim 55$ K and does not show any obvious anomaly associated with the formation of the superconducting state.  For CaK(Fe$_{0.967}$Ni$_{0.033}$)$_4$As$_4$, $B_{hf}$ initially increases on cooling below $\sim 45$ K, and then, on further cooling below $T_c \approx 20$ K, decreases continuously. The theoretical discussion of this behavior is presented in the next section. This behavior is comparable to that observed in M\"ossbauer study of Ba$_{0.75}$K$_{0.25}$Fe$_2$As$_2$, \cite{mun13a} and in elastic neutron scattering data for transition metal substituted BaFe$_2$As$_2$, \cite{pra09a,chr09a,wan10a,luo12a} and recently  CaK(Fe$_{1-x}$Ni$_x$)$_4$As$_4$. \cite{kre18a}

Temperature and Ni-concentration dependences of the isomer shift and quadrupole splitting are presented in Appendix B.  Comparison of the temperature dependent, $^{57}$Fe hyperfine field with the temperature dependence of the ordered moment inferred from elastic neutron scattering is presented in Appendix C.

\section{Discussion}

\subsection{Suppression of magnetic order by the emerging superconducting state}

The problem of superconductivity coexisting with charge density wave order has been  considered  by  Bilbro and McMillan within a weak-coupling BCS  model for both order parameters.  \cite{bil76a}  K. Machida applied the same formalism to the question of coexistence of superconductivity and spin density wave. \cite{mac81a}  The model was developed for an anisotropic, three-dimensional, single band case, yet it captures the main experimental features.The superconducting critical temperature, in absence of magnetism, is given by  $\Delta_0(0)/k_B T_{c0}=\pi/e^C\approx1.76$ ($C\approx 0.577$ is the Euler constant, $\Delta_0(0)$ is the gap at $T=0$.). Similarly, for the pure magnetic order parameter we have $M_0(0)/k_B T_{s0}=\pi/e^C\approx1.76$; here $M_0$ is the energy gap in the electron spectrum over the salient part of the Fermi surface in the absence of superconducting order and  the transition temperature for the magnetic transition is $T_{s0} > T_{c0}$. The spin density wave (SDW) order is assumed to develop over a nested part of the Fermi surface with the relative density of states $N_1/N_0=n_1 < 1$, whereas the superconductivity forms over, and gaps the full Fermi surface with the DOS $N_0$ without SDW, and part of the DOS, $N_2 = N_0 - N_1$ when SDW is present. When both orders coexist, the order parameters $M(T)$ and $\Delta(T)$
satisfy the system of two coupled self-consistency equations \cite{mac81a}:
\begin{align}
 & \ln\frac{T}{T_{s0}}~=~2\pi T\sum_{\omega>0}^{\omega_s}\left[\frac{1}{2 M}\left( \frac{M+\Delta}{\sqrt{\omega^2+(M+\Delta)^2}}+ \frac{M-\Delta}{\sqrt{\omega^2+(M-\Delta)^2}}\right)-\frac{1}{\omega}\right]\,, \label{e1}\\
 & \ln\frac{T}{T_{c0}}~=~n_1 2\pi T\sum_{\omega>0}^{\omega_D}\left[\frac{1}{2 \Delta}\left( \frac{ \Delta+M}{\sqrt{\omega^2+( \Delta+M)^2}}+ \frac{ \Delta-M}{\sqrt{\omega^2+( \Delta-M)^2}}\right)- \frac{1}{\omega}\right] \nonumber \\
& + n_2 2\pi T\sum_{\omega}^{\omega_D}\left(\frac{1}{\sqrt{\omega^2+ \Delta^2}}-\frac{1}{\omega}\right) . \qquad \label{e2}
\end{align}
 Here, 
$\omega=\pi T(2n+1)$ are Matsubara frequencies with integer $n \geq 0$, $\omega_D$ is the Debye frequency,   $\omega_s$ is a corresponding limit for SDW, and $n_2=1-n_1$. For brevity we use units with Plank's $\hbar$ and Boltzmann's $k_B$ as unities,  so that temperature and frequency have units of energy. The sums here are convergent and for $\omega_D\gg T_{c0}$ and $\omega_s\gg T_{s0}$  the upper limits of summation can be extended to infinity. 
 
For numerical work aimed at the situation with $T_{s0}>T_{c0}$,   it is convenient to introduce dimensionless variables
\begin{align}
& t=\frac{T}{T_{s0}}\,,\quad  d=\frac{\Delta}{2 \pi T_{s0}}\,,\quad m= \frac{M }{2 \pi T_{s0}}\,. \label{d,m} 
\end{align}
After some rearrangements,   Eqs.\,(\ref{e1}), (\ref{e2}) take  the form:

\begin{align}
& m\ln t= \sum_{n \geq 0}^ \infty \left[\frac{t}{2  }\left( \frac{ m+d}{\sqrt{t^2(n+1/2)^2+(m+d)^2 }}+ \frac{m-d }{\sqrt{t^2(n+1/2)^2+(m-d)^2 }}\right)
-\frac{m}{n+1/2}\right] \,, \label{e4}\\
& d \ln (R\, t) = n_1 \sum_{n \geq 0}^ \infty \left[\frac{t}{2 }\left( \frac{ d+m}{\sqrt{t^2(n+1/2)^2+(m+d)^2 }}+ \frac{ d-m}{\sqrt{t^2(n+1/2)^2+(d-m )^2 }}\right)- \frac{d}{n+1/2}\right] \nonumber\\
 & + n_2 d  \sum_{n \geq 0}^{\infty}\left(\frac{t }{\sqrt{t^2(n+1/2)^2+ d ^2 }}-\frac{1}{n+1/2}\right) , \label{e5}
\end{align}

\noindent where $R=T_{s0}/T_{c0}>1$.  Fig.\ref{fig1} shows  numerical solutions for $n_1=0.05$ and $n_1=0.3$,  $R=T_{s0}/T_{c0}=2$ and $R = 4$.  Clearly, the SDW order parameter at $T_c < T < T_{s0}$ has a standard BCS temperature dependence. 

Fig. \ref{fig1} shows that the effect of superconductivity on the magnetic order parameter is larger for smaller values of $n_1$, e.g. for smaller nesting (for constant $R$), and for smaller $R$ (for constant $n_1$). Qualitatively, and expectedly, it means that (within the model) magnetism is more robust than superconductivity. To observe measurable suppression of magnetic order parameter below $T_c$ one has to have small nesting and/or not very different bare $T_{s0}$ and $T_{c0}$ values. Fig. \ref{fig1} also shows that one can have similar behavior of $m$ and $d$ as a function of temperature for different values of $R$ and $n_1$. As such, a unique determination $R$ and $n_1$ would require additional boundary conditions on them. 

To obtain an equation for $T_c$, the superconducting transition temperature in the presence of magnetic order, one multiplies Eq.\,(\ref{e5}) by $d$ and goes to the limit $d\to 0$:
\begin{align}
\ln (R t_c)=n_1 \sum_{n>0}^ \infty  \left( \frac{ (n+1/2)^2}{ [(n+1/2)^2+ m_c^2/ t_c ^2]^{3/2}} - \frac{1}{n+1/2}\right)   , 
\qquad \label{e6}
\end{align}
where $t_c=T_c/T_{s0}$ and $m_c$ is the normalized magnetization at $t_c$. This equation  contains two unknowns, $t_c $ and $m_c$. Since $d=0$ at $t_c$, the magnetization satisfies the equation for  $m_c(t_c)$:
\begin{align}
\ln  t_c= \sum_{n>0}^ \infty  \left( \frac{1}{ \sqrt{(n+1/2)^2+ m_c^2/ t_c ^2} } - \frac{1}{n+1/2}\right) . 
\qquad \label{e7}
\end{align}
In other words, for given $R$ and $n_1$, the system of Eqs.\,(\ref{e6}) and (\ref{e7}) can be solved for $t_c $ and $m_c$. The result is shown in Fig. \ref{fig2} for $R=2$; in particular, it shows that the superconductivity is practically suppressed for $n_1>0.8$. Grossly speaking, Fig. \ref{fig2} is an illustration of the fact that both the SDW and superconductivity are built from gapping Fermi surface; if there is almost no Fermi surface left for superconductivity, then $t_c$ drops toward zero.

 Figure\,\ref{fig3} shows that the experimental data for the two samples of CaK(Fe$_{1-x}$Ni$_x$)$_4$As$_4$ (magnetic hyperfine field serves as a proxy for magnetization) can be fit quite well by Machida's model. As discussed above, this is not necessarily a unique fit, and additional analysis and experimental data are required to justify these particular values of model parameters. It is important to stress that our measurements provide direct access to  magnetic order parameter magnitude, not just usually measured transition temperature. Thus we demonstrate directly that superconductivity does suppress the spin density wave order, in agreement with the theoretical analysis.

\subsection{Magnetic hyperfine field,  N\'eel temperature and x - T phase diagram}

Analysis of the experimental data of magnetic hyperfine field and the magnetic ordering temperature (see e.g. Ref. [\onlinecite{gol14a}]) suggested proportionality between $B_{hf}$ at base temperature and $T_N$ that translates into $T_N \propto M$, where M is the Fe effective moment. For CaK(Fe$_{0.951}$Ni$_{0.049}$)$_4$As$_4$ superconductivity has no apparent effect on $B_{hf}(T)$ (Fig. \ref{fig3}).  To evaluate the hyperfine field at base temperature in absence of superconductivity for For CaK(Fe$_{0.967}$Ni$_{0.033}$)$_4$As$_4$  we use the results of fits in Fig. \ref{fig3}.

The plot of $B_{hf}$ vs $T_N$ for these two compounds together with the literature data for several members of 122 and 1111 families is shown in Fig. \ref{TN}. Although, for the two 1144 compounds studied here, the difference between the values of $T_N$ and the inferred values of $B_{hf}$ is rather small, it appears that the gross trend of $B_{hf} \propto T_N$ observed in 122 family probably holds for 1144, although studies on larger set of samples are required to support (or refute) this statement.

Finally, the thermodynamic, specific heat (Appendix A), and spectroscopic, M\"ossbauer, measurements allow us to confirm and refine the $x - T$ phase diagram for CaK(Fe$_{1-x}$Ni$_x$)$_4$As$_4$. \cite{mei18a} Both experimental techniques used in this work allow for the detection of  magnetic ordering above, as well as below, the superconducting transition. For the $x$ = 0.017 sample there is no broadening of the M\"ossbauer spectra at low temperatures, that could be associated with a static magnetic hyperfine field on the $^{57}$Fe site and no additional anomalies in $C_p(T)$ below $T_c$. Consequently no long range magnetic order exists for CaK(Fe$_{0.983}$Ni$_{0.017}$)$_4$As$_4$, at least above either 5.5 K ($B_{hf} = 0$) or 1.9 K ($C_p(T)$). The current suggested $x - T$ phase diagram is shown in Fig. \ref{PD}. This phase diagram is consistent with the rather general, simple model in Ref. [\onlinecite{mac81a}] that predicts that the magnetic spin density wave state is precluded when the superconductivity develops at a higher temperature, since the superconducting energy gap opens all over the Fermi surface and prohibits the formation of the spin density wave gap.   On the other hand, when the onset temperature of the spin density wave is higher than that of superconductivity, these two long range orders, according to  Ref. [\onlinecite{mac81a}]  generally coexist. It is noteworthy that recent theoretical work on coexistence of superconductivity and magnetism in iron pnictides \cite{fer10a} suggested similar $x - T$ phase diagram for the case of $s^{ \pm}$ superconducting pairing. Further studies for $0.017 < x < 0.033$ will be needed to determine fine details of whether there is "back-bending" of the $T_N$ line once $T_N$ drops below $T_c$.

\section{Summary}

Our $^{57}$Fe M\"ossbauer study of  CaK(Fe$_{1-x}$Ni$_x$)$_4$As$_4$ compounds detected no magnetic hyperfine field (e.g. no static magnetic order) down to 5.5 K for $x$ = 0.017 and followed the evolution of $B_{hf}$ with temperature for $x$ = 0.033 and 0.049. The long-range magnetic spin-vortex-crystal order \cite{mei18a} was found to coexist with superconductivity, however, similar to the doped 122 compounds, the magnetic hyperfine field (ordered magnetic moment) below $T_c$ in  CaK(Fe$_{0.967}$Ni$_{0.033}$)$_4$As$_4$  is continuously suppressed with the developing superconducting order parameter. The $B_{hf}(T)$ data for  CaK(Fe$_{0.967}$Ni$_{0.033}$)$_4$As$_4$, and  CaK(Fe$_{0.951}$Ni$_{0.049}$)$_4$As$_4$ were analyzed using the  model of Machida for coexistence of itinerant spin density wave magnetism and superconductivity. \cite{mac81a} It is remarkable that this rather simple model can account for experimental observations in real, complex materials.

Similarly to 122 compounds, the values of $T_N$ and base temperature $B_{hf}$ are roughly proportional, suggesting that the value of $T_N$ in the  CaK(Fe$_{1-x}$Ni$_x$)$_4$As$_4$ family is mainly affected by the value of the magnetic moment on iron. 

In addition, specific heat data on  CaK(Fe$_{1-x}$Ni$_x$)$_4$As$_4$ (Appendix A) allowed for additional thermodynamically determined points on the $x - T$ phase diagram as well as additional values of $\Delta C_p$ at $T_c$ which were found to follow  BNC scaling. \cite{bud09a}

The isomer shift was found to  have insignificant Ni-concentration dependence, whereas both quadrupole splitting and line width monotonically increase  with Ni concentration.

\begin{acknowledgments}

Work at the Ames Laboratory was supported by the U.S. Department of Energy, Office of Science, Basic Energy Sciences, Materials Sciences and Engineering Division. The Ames Laboratory is operated for the U.S. Department of Energy by Iowa State University under contract No. DE-AC02-07CH11358. WRM was supported by the Gordon and Betty Moore Foundation's
EPiQS Initiative through Grant GBMF4411.

\end{acknowledgments}

\appendix
\section{Specific heat}

In addition to electrical resistivity and magnetic susceptibility measurements \cite{mei16a,mei18a} on the CaK(Fe$_{1-x}$Ni$_x$)$_4$As$_4$ samples with $x$ =0,  0.017, 0.033, and 0.049, the temperature dependent specific heat measurements, using a hybrid adiabatic relaxation technique of the heat capacity option in a Quantum Design, Physical Property Measurement System instrument were performed on these samples. The data, plotted as $C_p/T$ vs $T$ are shown in Fig. \ref{FA1}

The data clearly show the evolution of the superconducting and magnetic transitions with Ni- substitution. $T_c$ decreases with Ni-doping, in agreement with the published phase diagram \cite{mei18a} as does the jump  in the specific heat at $T_c$. The signatures corresponding to the magnetic phase transitions are observed only for $x$ = 0.033, 0.049, with no anomaly below $T_c$ found for $x = 0$ or $x$ = 0.017. Altogether the specific heat data allows to confirm and refine, with a thermodynamic measurement, the $x - T$ phase diagram  for CaK(Fe$_{1-x}$Ni$_x$)$_4$As$_4$ suggested in Ref. [\onlinecite{mei18a}].

It has been shown \cite{bud09a,kim11a,kim12a,bud15a,ban17a} that for many iron-based superconductors, in particular of 122 family, an empirical trend, so called BNC scaling, $\Delta C_p|_{T_c} \propto T_c^3$  is observed. Moreover, deviation from such scaling was suggested to be a signature of significant changes in the nature of the superconducting state. \cite{bud15a,bud13a,gri14a}  The data for CaK(Fe$_{1-x}$Ni$_x$)$_4$As$_4$ ($x$ =0,  0.017, 0.033, and 0.049) were added to the BNC plot (Fig. \ref{BNC}) (to be consistent with the previous data for the 122 family, for this plot the molecular weight was taken as 1/2 of the molecular weight of CaK(Fe$_{1-x}$Ni$_x$)$_4$As$_4$). These data agree well with the rough,  $\Delta C_p|_{T_c} \propto T_c^3$ trend, suggesting that the nature of superconductivity is probably similar to that in the majority of the members of the 122 family. At the same time there data are consistent with superconductivity in CaK(Fe$_{1-x}$Ni$_x$)$_4$As$_4$ being bulk.

\section{Hyperfine parameters}

Isomer shift and quadrupole splitting as a function of temperature are plotted for  CaK(Fe$_{1-x}$Ni$_x$)$_4$As$_4$, $x$ = 0, 0.017, 0.033, and 0.049 in Fig. \ref{FA3}.  Taken together, all data are very consistent. In the paramagnetic state the isomer shift for all four compounds is almost the same (it decreases by $\sim 2\%$ between $x$ = 0 and $x$ = 0.049, Fig. \ref{HF}). This means that the changes in the local electron density at the iron site, as well as the difference in the Debye temperatures that dominate the $IS(T)$ dependence, are insignificant (cf. small $< 4\%$ changes in the IS values in the (Ba$_{1-x}$K$_x$)(Fe$_{1-y}$Co$_y$)$_2$As$_2$ \cite{gol14a}). The quadrupole splitting increases with Ni - substitution (Fig. \ref{HF}). This could be related to the change of local environment of the $^{57}$Fe accompanying  change of the lattice parameters (see Ref. [\onlinecite{mei18a}], Supplemental Information), however further structural work as well as band structure calculations would be required to understand this trend. 

For CaK(Fe$_{1-x}$Ni$_x$)$_4$As$_4$, $x$ = 0.033, and 0.049 there is minor change in the isomer shift values between paramagnetic and the magnetically ordered state. The increase of $IS$ by $\sim 5\%$ suggests that the local electron density at the iron site increases in the magnetically ordered state. Some changes of electronic structure in the ordered state are expected, since the magnetic unit cell doubles in in the spin-vortex-crystal state. ARPES experiments are desirable for understanding of these changes. The is no apparent change in the $\lvert QS \rvert$ at the transition  within the scattering of the results.

\section{Comparison with neutron scattering data}

Temperature dependent, hyperfine field data for CaK(Fe$_{0.967}$Ni$_{0.033}$)$_4$As$_4$, and  CaK(Fe$_{0.951}$Ni$_{0.049}$)$_4$As$_4$ are plotted in Fig. \ref{Fcompare} together with  the square root of the intensity measured at the (1/2~1/2~3) antiferromagnetic Bragg peak position for both samples that is proportional to the antiferromagnetic moment, the antiferromagnetic order parameter. \cite{kre18a} These two sets of data scale fairly well, with scaling coefficient being different by $\sim 12$\% between $x$ = 0.033 and 0.049 data sets. This comparison of two data sets, obtained on the samples grown in very similar way, give confidence in use of M\"ossbauer spectroscopy for further studies of coexistence of superconductivity and magnetism in iron-based superconductors. In addition, this comparison allows for the evaluation of the ratio between the magnetic hyperfine field and the magnetic moment ($A$) in the 1144 materials. \cite{dub09a} Taking two values of the magnetic moment cited in Ref. [\onlinecite{kre18a}] and comparing them with the corresponding values of $B_{hf}$ yields $A \approx 6.3$ T/$\mu_B$. This value is the same as repoted for  BaFe$_2$As$_2$ \cite{rot08a,hua08a}.

\clearpage

\begin{figure}
\begin{center}
\includegraphics[angle=0,width=140mm]{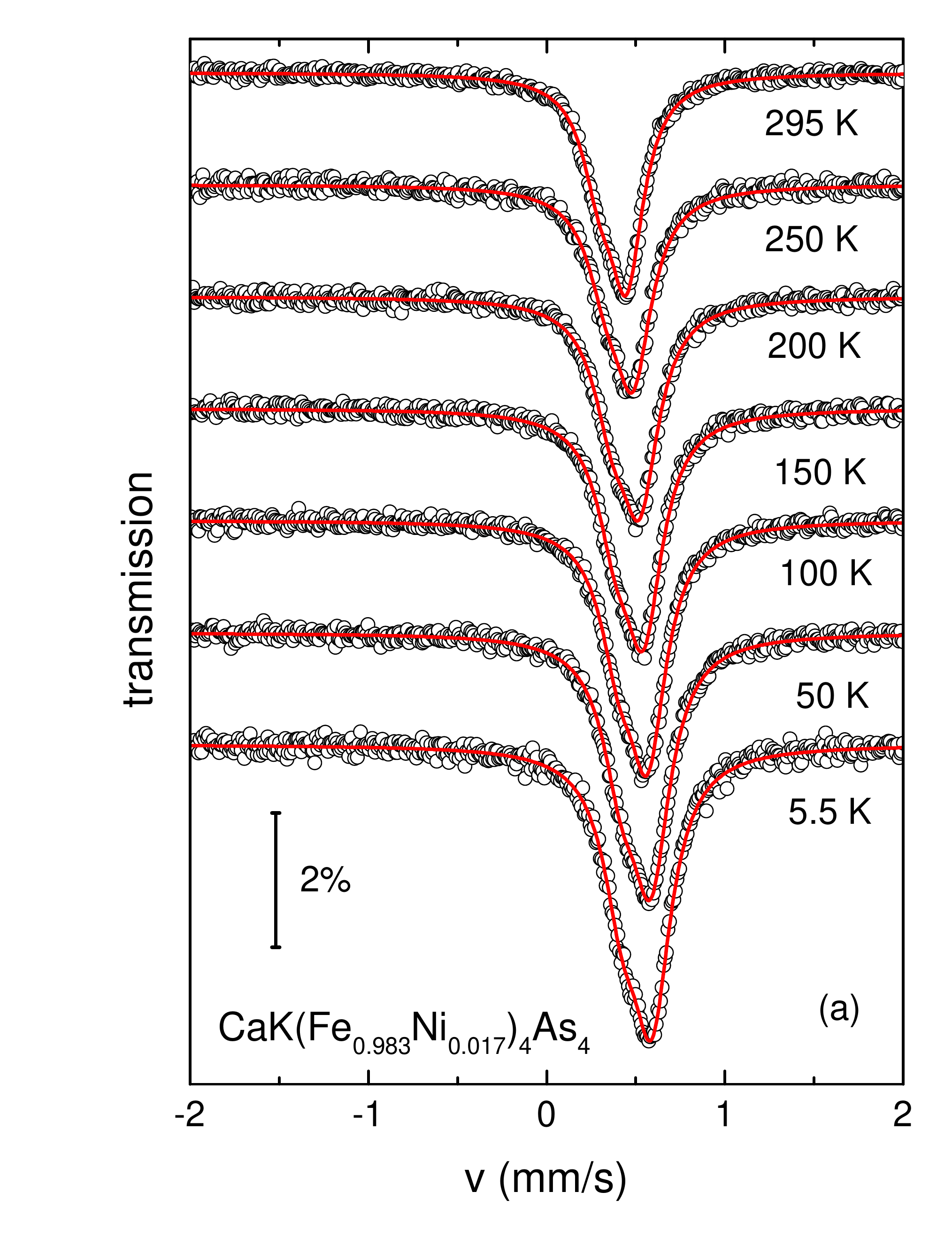}
\end{center}
\end{figure}

\clearpage

\begin{figure}
\begin{center}
\includegraphics[angle=0,width=140mm]{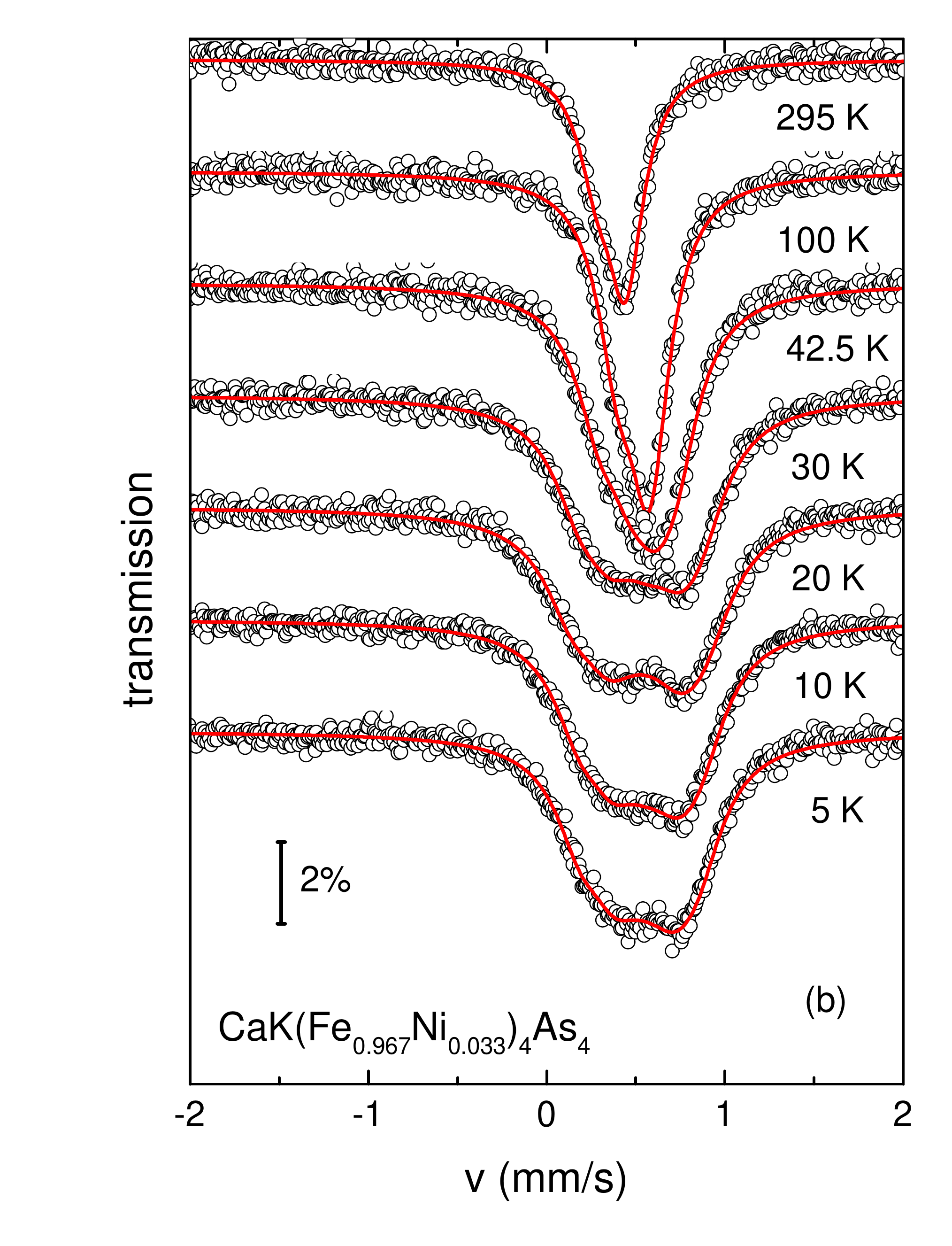}
\end{center}
\end{figure}

\clearpage

\begin{figure}
\begin{center}
\includegraphics[angle=0,width=140mm]{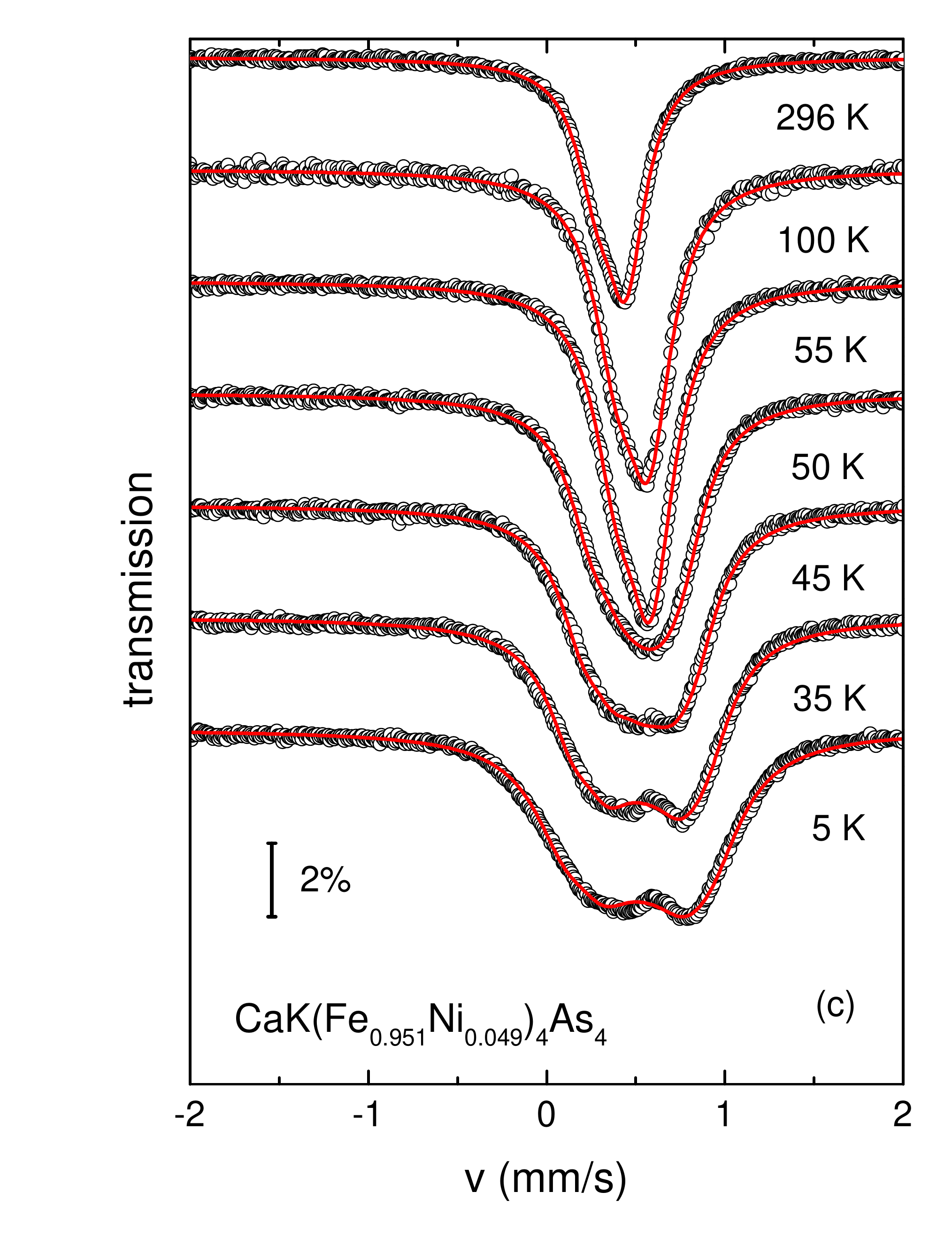}
\end{center}
\caption{(color online) $^{57}$Fe M\"ossbauer spectra of  (a) CaK(Fe$_{0.983}$Ni$_{0.017}$)$_4$As$_4$,  (b) CaK(Fe$_{0.967}$Ni$_{0.033}$)$_4$As$_4$, and  (c) CaK(Fe$_{0.951}$Ni$_{0.049}$)$_4$As$_4$,   at selected temperatures. Symbols - data, lines - fits.} \label{F1}
\end{figure}

\clearpage

\begin{figure}  
\begin{center}
\includegraphics[width=120mm]{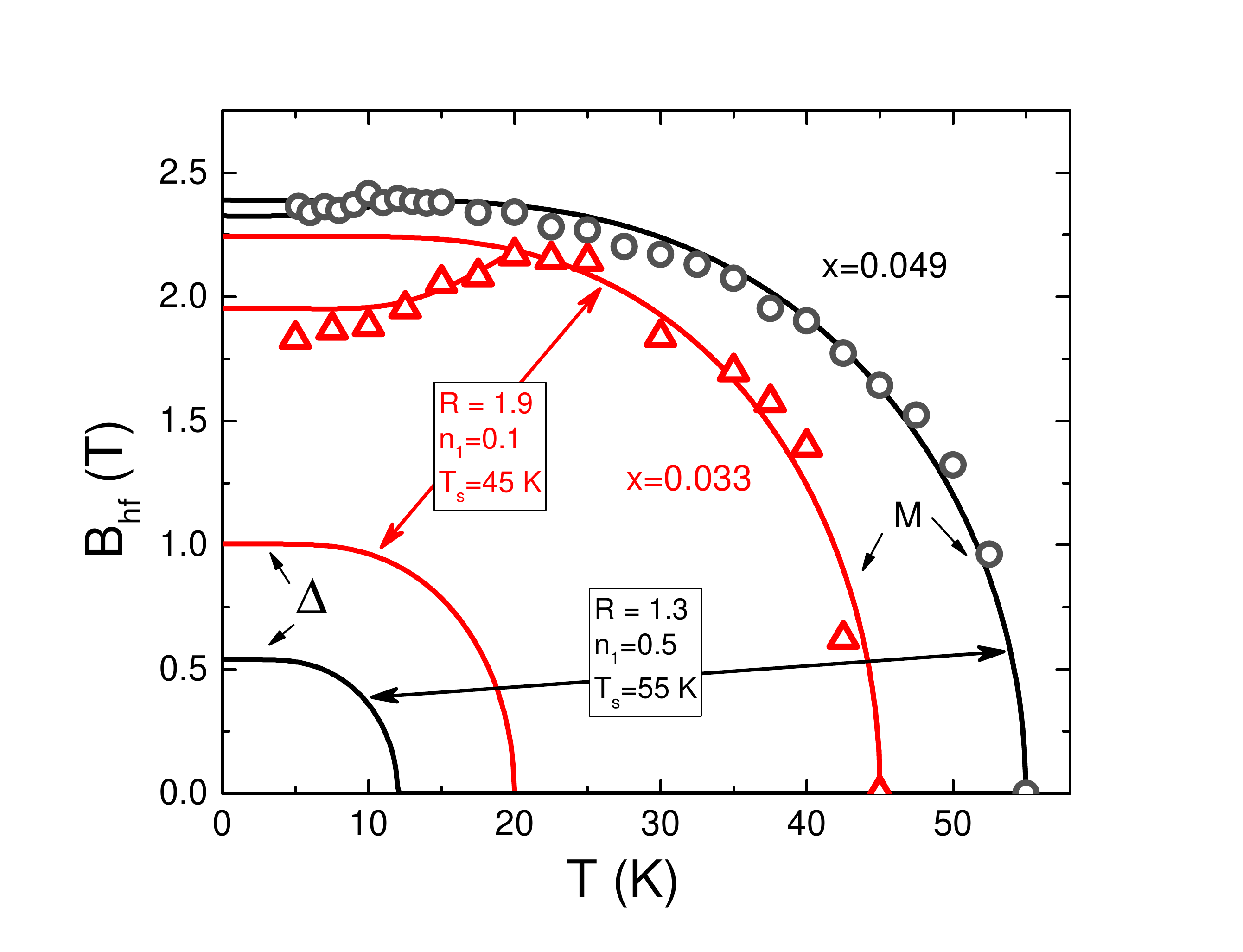}
\end{center}
\caption{ (Color online) Experimental data (symbols) of $B_{hf}(T)$ for CaK(Fe$_{0.967}$Ni$_{0.033}$)$_4$As$_4$, and  CaK(Fe$_{0.951}$Ni$_{0.049}$)$_4$As$_4$  overlayed with temperature dependence of scaled magnetic, M, and superconducting, $\Delta$ order parameters (lines) from fits using model of Ref. [\onlinecite{mac81a}] (with $B_{hf}(T)$ serving as a proxy for magnetization). Obtained fitting parameters are listed on the plot.}
\label{fig3}
\end{figure}

\clearpage

\begin{figure}
\begin{center}
\includegraphics[width=120mm]{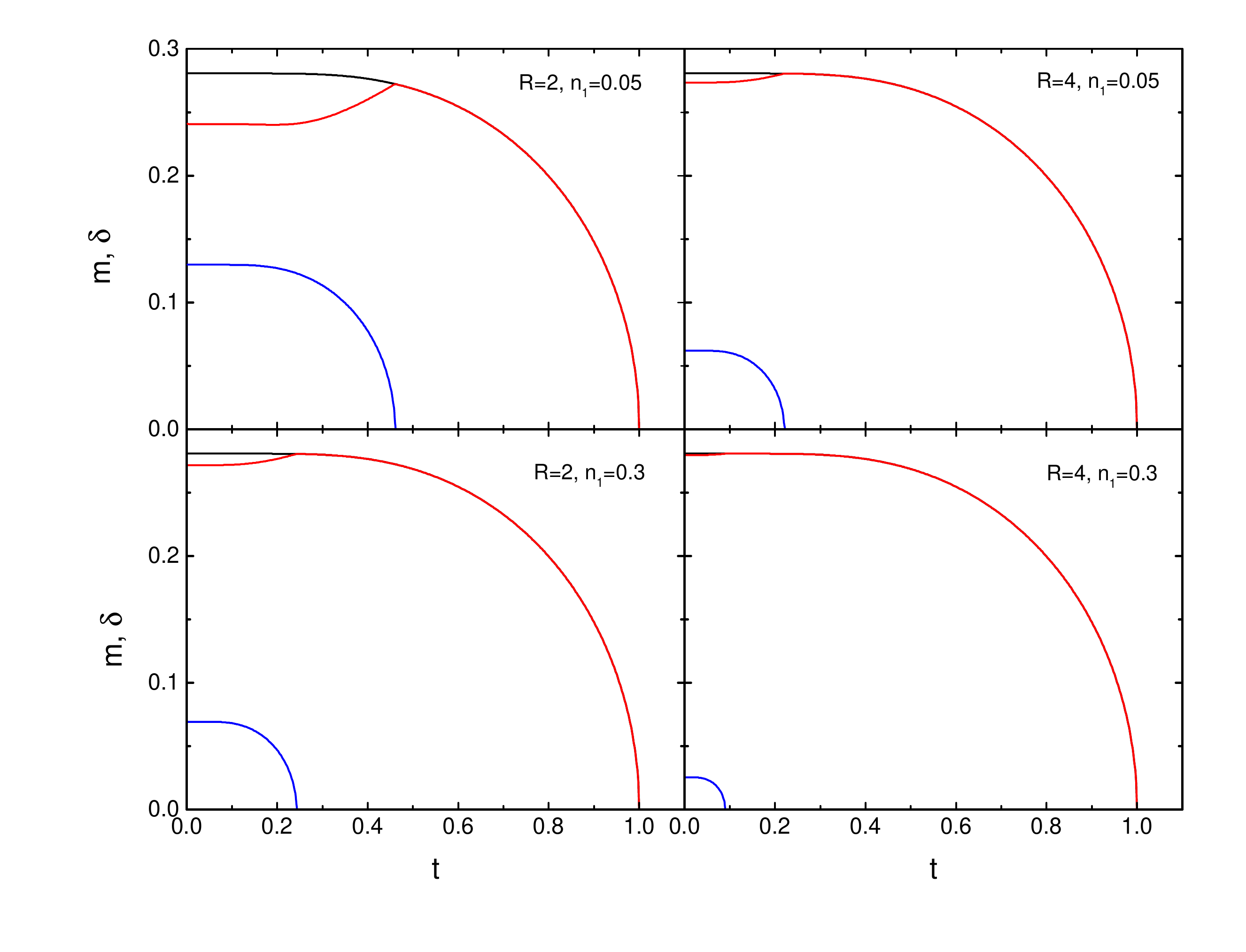}
\end{center}
\caption{ (Color online) Magnetic (red line) and superconducting (blue line) order parameters  for $R=T_{s0}/T_{c0}=2$ and $R = 4$. The black line is for part of  ``bare" $m_0(t)$ below $T_c$. (See text for details)}
\label{fig1}
\end{figure}

\clearpage

\begin{figure}
\begin{center}
\includegraphics[width=120mm]{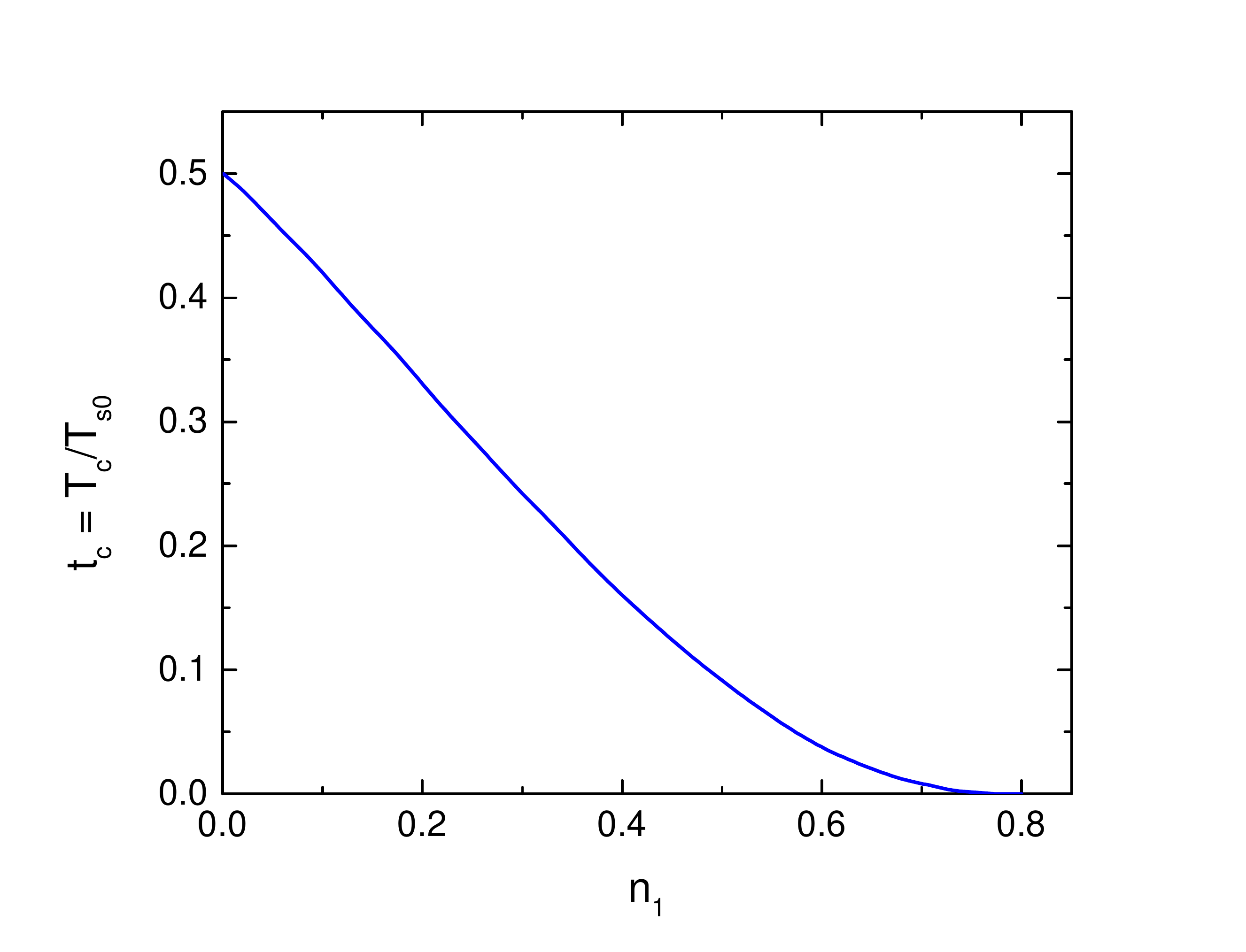}
\end{center}
\caption{ (Color online) $t_c=T_c/T_{s0}$ as a function of $n_1$ for $R=2$.}
\label{fig2}
\end{figure}

\clearpage

\begin{figure}
\begin{center}
\includegraphics[angle=0,width=120mm]{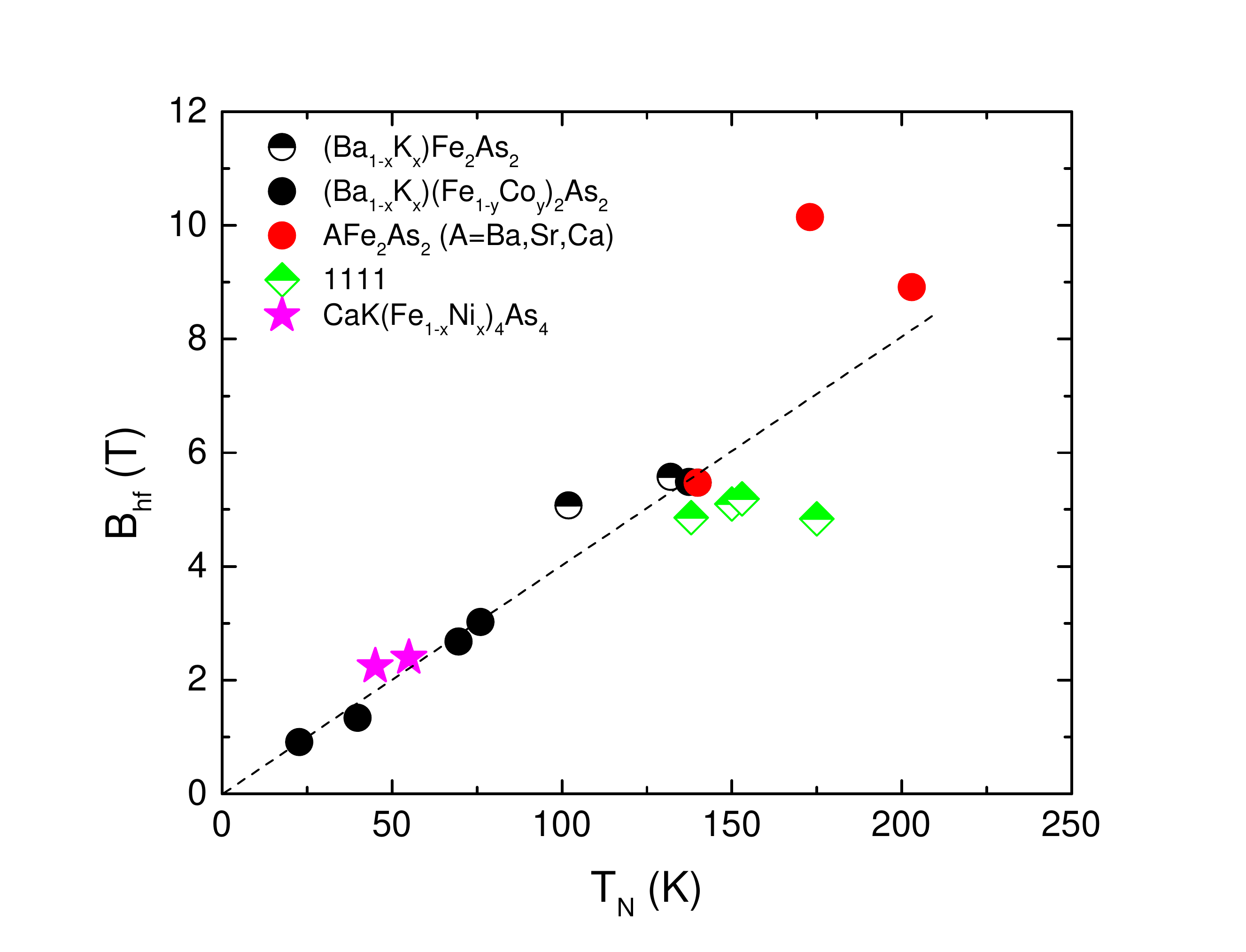}
\end{center}
\caption{(color online) The value of magnetic hyperfine field on $^{57}Fe$ sites as a function of the N\'eel temperature for several Fe-based superconductors and related materials. Data for    CaK(Fe$_{1-x}$Ni$_x$)$_4$As$_4$ - this work, other points are taken from the literature. \cite{gol14a,rot08a,rot09a,teg08a,alz11a,now08a,kla08a,mcg08a,teg08b} Dashed line - linear fit for (Ba$_{1-x}$K$_x$)(Fe$_{1-y}$Co$_y$)$_2$As$_2$. \cite{gol14a}} \label{TN}
\end{figure}

\clearpage

\begin{figure}
\begin{center}
\includegraphics[angle=0,width=120mm]{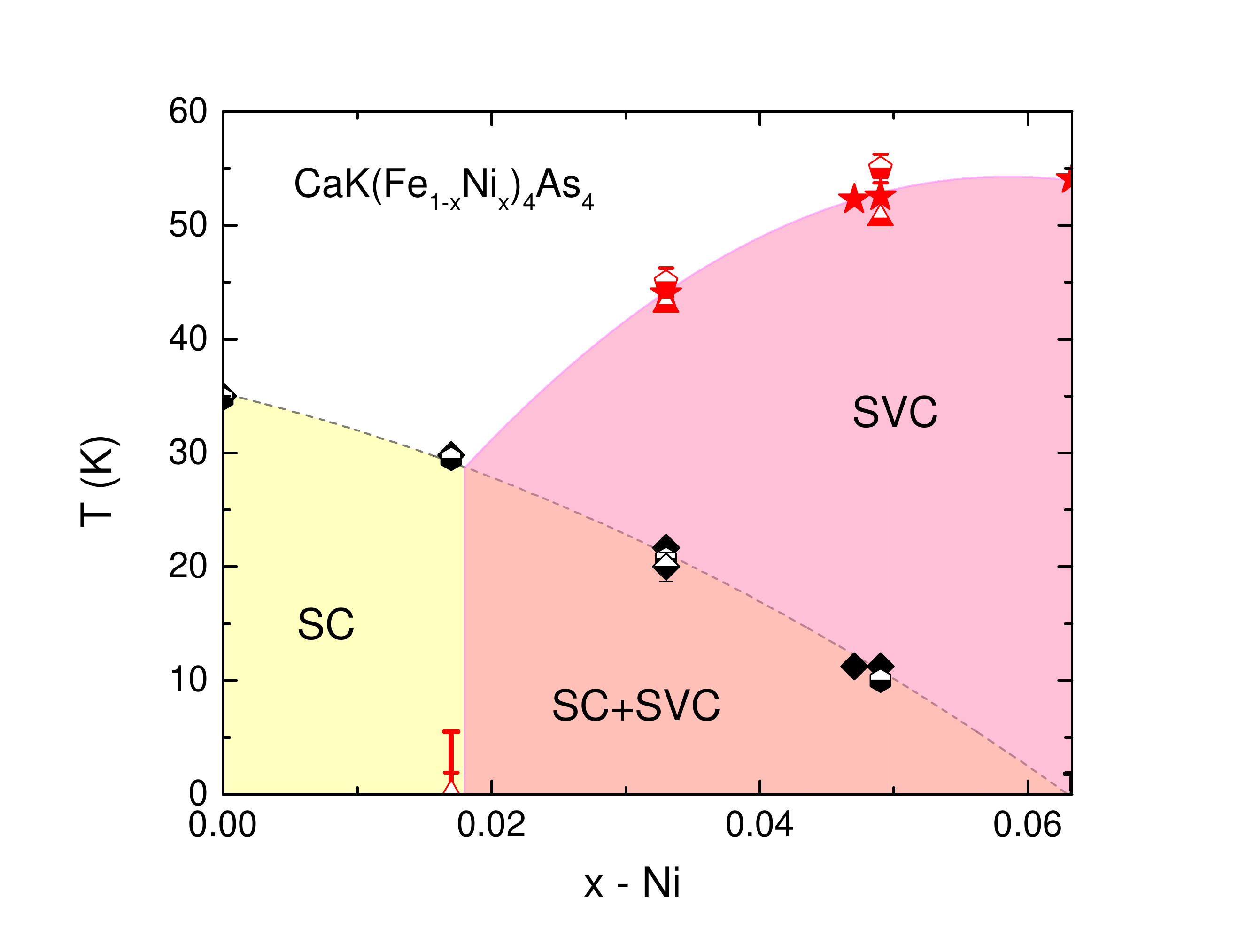}
\end{center}
\caption{(color online) $x - T$ phase diagram for  CaK(Fe$_{1-x}$Ni$_x$)$_4$As$_4$. Phases: SC - superconducting; SVC - magnetic spin vortex crystal, SC+SVC - coexistence of superconductivity and spin vortex crystal magnetic order. Symbols: filled - from Ref. [\onlinecite{mei18a}], half - filled - this work: triangles and hexagons - from $C_p(T)$, pentagons and rhombus - from M\"ossbauer spectroscopy. Lines are guides for the eye.} \label{PD}
\end{figure}

\clearpage

\begin{figure}
\begin{center}
\includegraphics[angle=0,width=120mm]{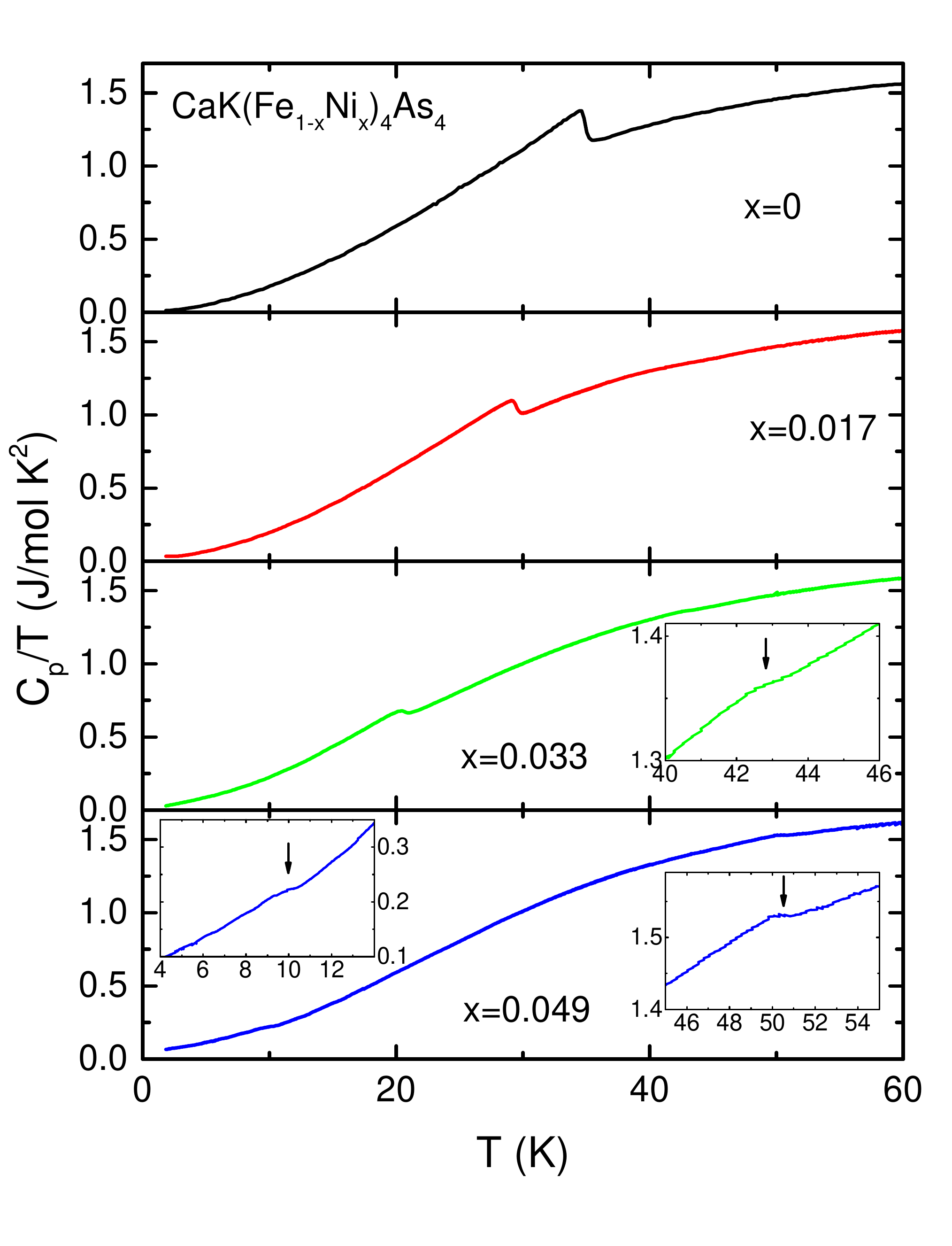}
\end{center}
\caption{(color online) Temperature dependent specific heat of CaK(Fe$_{1-x}$Ni$_x$)$_4$As$_4$ samples with $x$ = 0, 0.017, 0.033, and 0.049 plotted as $C_p/T$ vs $T$. Insets - enlarged parts of the plots at magnetic ($x$ = 0.033, 0.049 panels, right) and superconducting ($x$ = 0.49 panel, left) transitions. Arrows in the insets mark the transition temperatures. Some of the data were previously shown  in Refs. [\onlinecite{mei16a},\onlinecite{mei18a}].  } \label{FA1}
\end{figure}

\clearpage

\begin{figure}
\begin{center}
\includegraphics[angle=0,width=120mm]{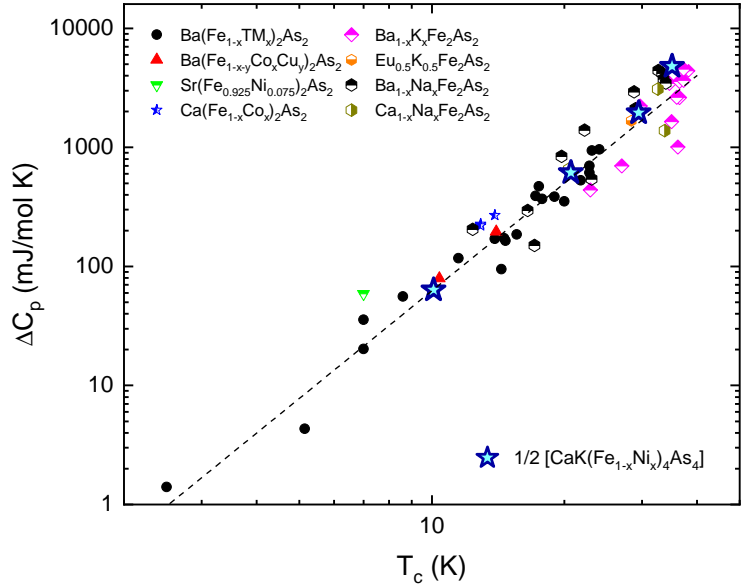}
\end{center}
\caption{(color online) $\Delta C_p$ at the superconducting transition vs $T_c$ for the  CaK(Fe$_{1-x}$Ni$_x$)$_4$As$_4$ samples with $x$ = 0, 0.017, 0.033, and 0.049 plotted together with literature data \cite{bud15a} for various Fe-based superconductors. For consistency, half of the molecular weight of 1144 samples was taken for this plot. Literature data for KFe$_2$As$_2$ and close concentrations \cite {bud15a} are not shown for simplicity.} \label{BNC}
\end{figure}

\clearpage

\begin{figure}
\begin{center}
\includegraphics[angle=0,width=120mm]{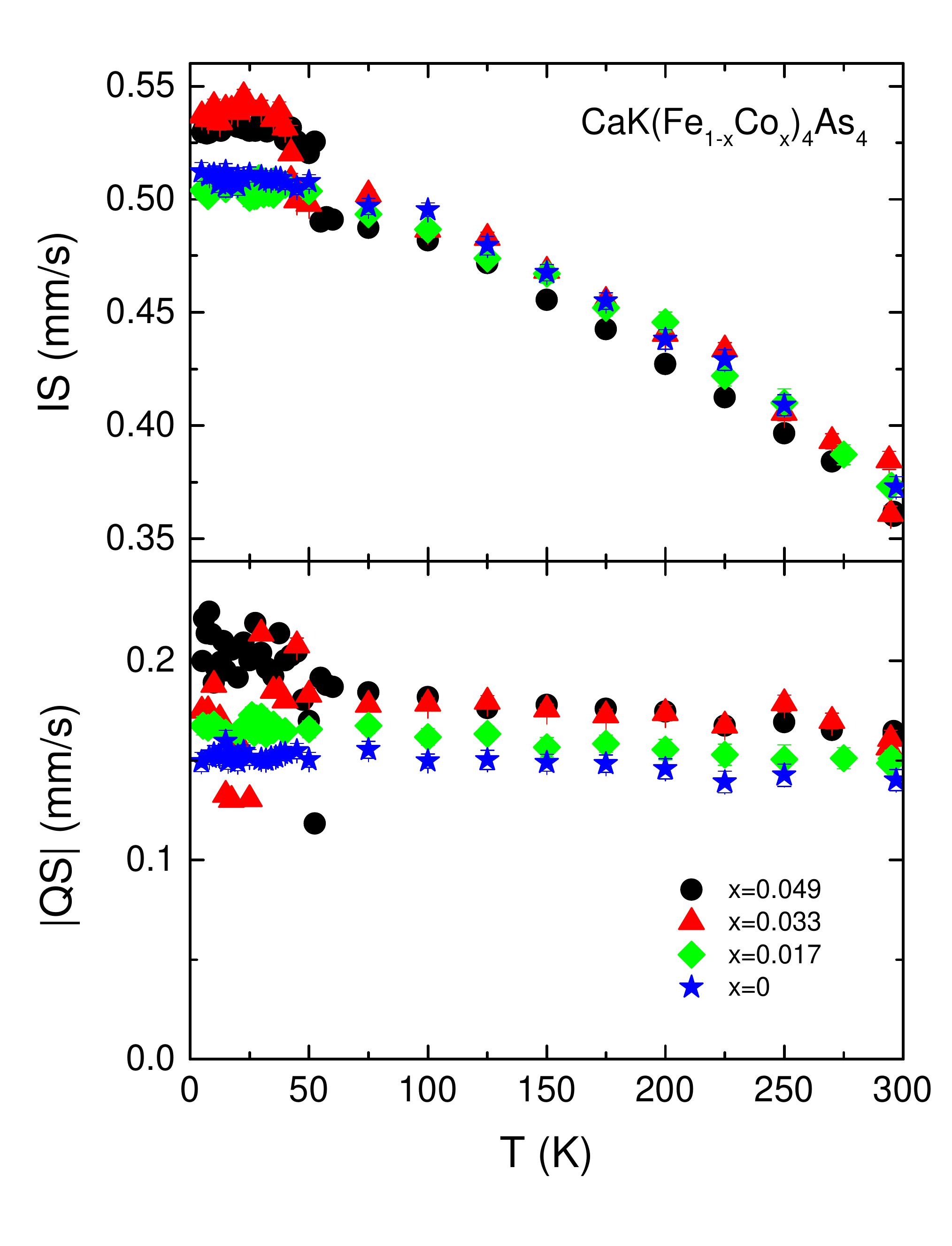}
\end{center}
\caption{(color online) Temperature dependent hyperfine parameters obtained from fits of  $^{57}$Fe M\"ossbauer spectra of CaK(Fe$_{1-x}$Ni$_x$)$_4$As$_4$ samples with $x$ = 0, 0.017, 0.033, and 0.049. at different temperatures: (a) isomer shift (IS), (b) absolute value of the quadrupole splitting ($\mid$QS$\mid$). Data for CaKFe$_4$As$_4$ are taken from Ref. [\onlinecite{bud17a}].} \label{FA3}
\end{figure}

\clearpage

\begin{figure}
\begin{center}
\includegraphics[angle=0,width=120mm]{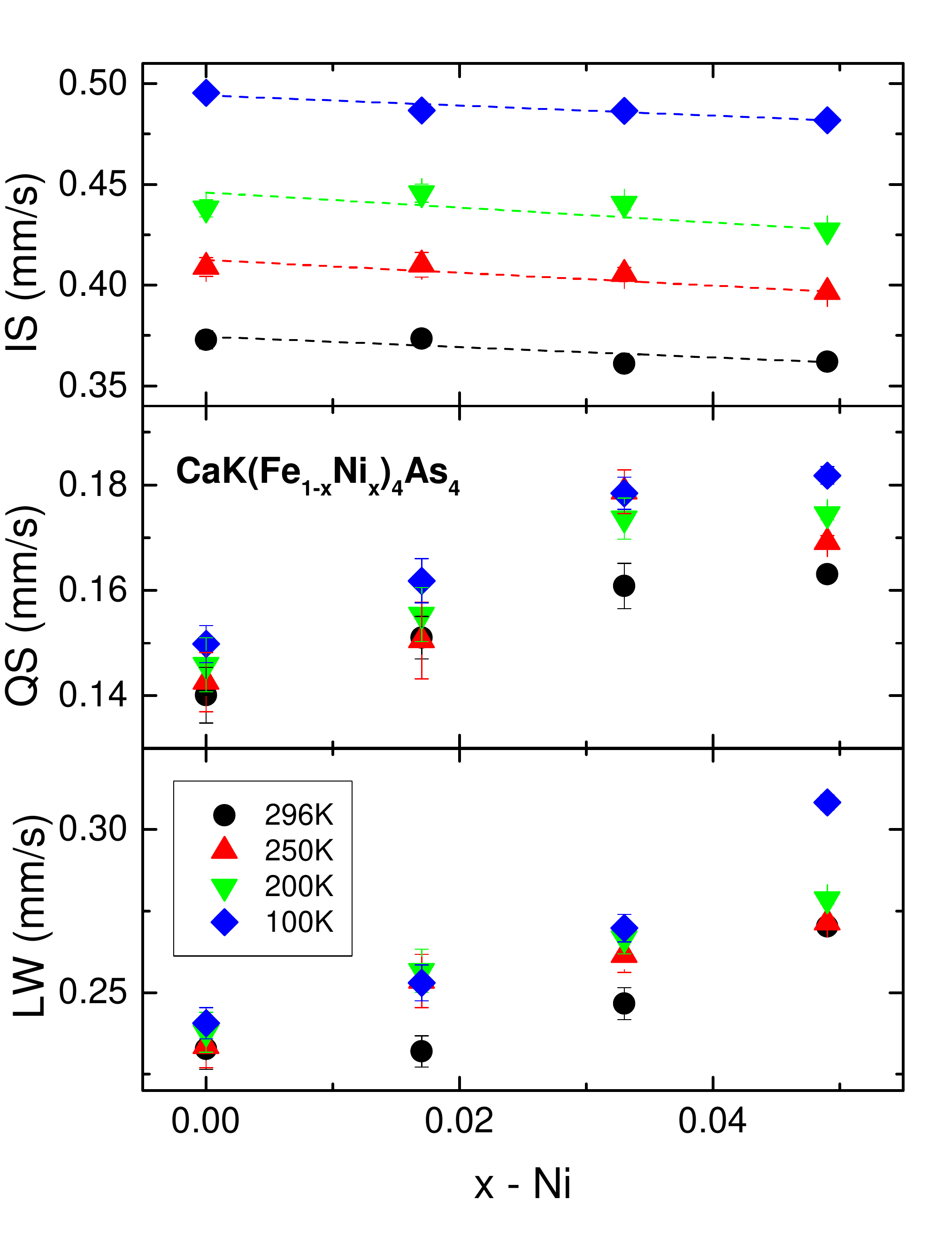}
\end{center}
\caption{(color online) Hyperfine parameters,  isomer shift, (b) quadrupole splitting, and line width at selected temperatures, obtained from fits of  $^{57}$Fe M\"ossbauer spectra of CaK(Fe$_{1-x}$Ni$_x$)$_4$As$_4$ samples with $x$ = 0, 0.017, 0.033, and 0.049, plotted as a function of Ni - concentration, $x$.  Data for CaKFe$_4$As$_4$ are taken from Ref. [\onlinecite{bud17a}]. Dashed lines - linear fits.} \label{HF}
\end{figure}

\clearpage

\begin{figure}
\begin{center}
\includegraphics[angle=0,width=120mm]{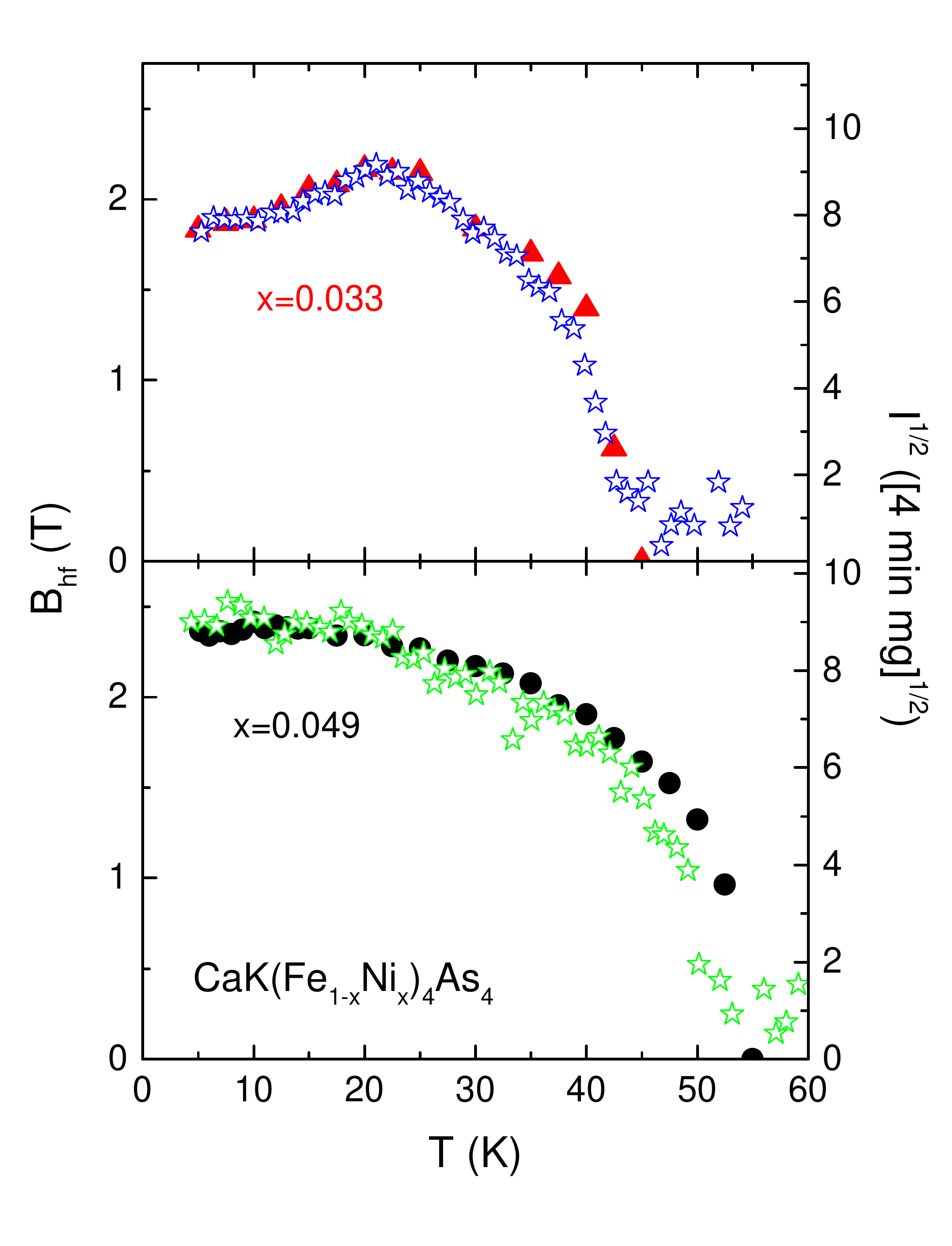}
\end{center}
\caption{(color online) Temperature dependent hyperfine field (filled and half-filled symbols, left axes) of CaK(Fe$_{1-x}$Ni$_x$)$_4$As$_4$ samples with $x$ =  0.033, and 0.049, plotted together with the square root of the intensity of the (1/2~1/2~3) antiferromagnetic Bragg peak from neutron scattering data in Ref. [\onlinecite{kre18a}] (open symbols, right axes).} \label{Fcompare}
\end{figure}

\end{document}